\documentclass{moriond}

\usepackage{natbib}
\usepackage{longtable} 
\usepackage{bm}

 \def\prl{Phys.~Rev.~Lett. }
 \def\aap{Astron. Astrophys. }
 \def\mnras{Mon. Not. R. Astron. Soc.}
 \def\prd{Phys.~Rev.~D }%
  \def\aj{Astron. J }%
  \def\ssr{Space~Sci.~Rev. }%
   \def\apj{ApJ }%
\def\lg{\lambda_g}

\def\i{\mathrm{i}}

\def\({\left(}
\def\){\right)}
\def\[{\left[}
\def\]{\right]}

\def\be{\begin{equation}}
\def\ee{\end{equation}}
\def\bea{\begin{eqnarray}}
\def\eea{\end{eqnarray}}

\newcommand{\Photo}{}

\begin{document}
\title{INPOP Planetary ephemerides and applications in the frame of the BepiColombo mission including new constraints on the graviton mass and dilaton parameters}

\author{ A. Fienga$^{1,2}$, L. Bernus $^{2,1}$ ,  O. Minazzoli$^{3}$,
   A. Hees$^{4}$,
   L. Bigot $^{6}$,
   C. Herrera $^{1}$,
 V. Mariani $^{1}$,
  A. Di Ruscio$^{5,1}$,
  D. Durante$^{5}$,
 D. Mary $^{6}$ 
}

\address{$^{1}$G\'eoazur, Observatoire de la C\^ote d'Azur, Universit\'e C\^ote d'Azur,CNRS, France \\  
$^{2}$IMCCE, Observatoire de Paris, PSL University, CNRS, France\\
 $^{3}$Artemis, Universit\'e C\^ote d'Azur, CNRS, Observatoire de la C\^ote d'Azur, France\\
 $^4$SYRTE,  Observatoire  de  Paris,  Universit\'e  PSL, CNRS,  Sorbonne  Universit\'e,  LNE,  France \\
 $^{5}$ CRAS, Sapienza University of Rome, Italy \\
 $^{6}$Lagrange, Observatoire de la C\^ote d'Azur, Universit\'e C\^ote d'Azur, CNRS, France
 }


\maketitle\abstracts{We present here the new results obtained with the INPOP planetary ephemerides and  BepiColombo radio-science simulations. We give new constraints for the classic General Relativity tests in terms of violation of the PPN parameters $\beta$ and $\gamma$ and the time variation of the gravitational constant G. We also present new limits for the mass of the graviton and finally we obtain new acceptable intervals for the dilaton parameters $\alpha_{0}$, $\alpha_{T}$ and $\alpha_{G}$. Besides these tests of gravitation, we also study the possibility of detecting the Sun core rotation.}

\section{INPOP planetary ephemerides and Bepi-Colombo simulations}

The INPOP20a planetary ephemerides has several modifications relative to INPOP19a \citep{2019NSTIM.109.....V} : the addition of 5 Jupiter positions deduced from the Juno perijove PJ19 to PJ23, leading to a coverage of more than 4 years of observations with an accuracy of about tens of meters on the Earth-Jovian barycenter measurements. Two important changes have also been brought to the dynamical modeling : first the introduction of the Lense-Thirring acceleration and second, a more realistic modeling of the accelerations induced by the Trans-Neptunian objects on the outer solar system. A full description of  the INPOP20a modele and fit can be found in \citep{2021arXiv211104499F,inpop21a}. INPOP21a  \citep{inpop21a} and INPOP22a are two updates of INPOP20a with more Juno and MEX data.

\subsection{Sun core rotation and Lense-Thirring effects}
 \label{sec:srot}
 
When comparing the estimations of the Sun oblateness, $J^{\odot}_{2}$, obtained with planetary ephemerides to values obtained by helio-seismology \citep{2008A&A...477..657A, 1998MNRAS.297L..76P}, it is important to keep in mind that an additional contribution must be included in order to compare consistent estimates: the effect of the Sun rotation on the space-time metric \citep{1918PhyZ...19..156L}. With the accuracy of the Bepi-Colombo mission, it is important to include the Lense-Thirring in the INPOP equations of motion as \cite{Hees2015} estimated that it contributes to 10$\%$ of the dynamical acceleration induced by the quadrupole of the Sun in General Relativity (GRT). The acceleration induced by the Lense-Thirring effect generated by a central body (at the first post-Newtonian approximation) is given by
\begin{equation}
\vec{a}_{LT} = \frac{(\gamma+1)G}{c^{2}r^{3}} S [ 3 \frac{\vec{k} . \vec{r}}{r^{2}} (\vec{r} \wedge \vec{v}) -  (\vec{k} \wedge \vec{v}) ]
\label{eq:LT}
\end{equation}
where $G$ is the gravitational constant, $c$ the speed of light, $\vec{S}$ is the Sun angular momentum such as $\vec{S} = S \vec{k}$ where $\vec{k}$ is the direction of the Sun rotation pole defined according to the IAU right ascension and declination \citep{2018CeMDA.130...22A}. $\vec{r}$ and $\vec{v}$ are the position and velocity vectors of the planet relative to the central body (here the Sun) and $\gamma$ is the PPN parameter for the light deflection.
The amplitude of the Sun angular momentum S, implemented in INPOP depends on the hypothesis for the Sun  rotation. As the rotation of the convective layers is well constrained by helio-seismology and composed about 80$\%$ of the total rotation rate \citep{2003ApJ...586..650K}, the remaining uncertainty on the Sun rotation comes from its core rotation \citep{2003LNP...599...31D}. For the very center of the Sun, the helioseismology based on acoustic modes is less precise. Several solutions presented in Table  \ref{tab:Sun} exist from a slow rotation rate \citep{1999MNRAS.308..405C} , an uniform rate at 435 nHz to a fast \citep{1998ESASP.418..329R} and a very fast rate \citep{2017A&A...604A..40F}. This very fast rotating core was suggested to explain a recent, but still debated, claim of g-mode detection. The way we derive the amplitude of Sun angular momentum S from the Sun core rotation is explained in  \citep{2021arXiv211104499F} and derived from the following equation :
\begin{equation}
S = \frac{1}{2} \int_0^{R_\odot} r^2 dm \int_{-1}^1 (1-\cos^2\theta)\,\Omega(r,\theta)\,d\cos\theta,
\label{eq:momemtum}
\end{equation}
 \\
where $\Omega(r,\theta)$ is the rotation rate of the sun that can be split into the contribution of the convective layers  given by helioseismology \cite{2001A&A...377..688R} and the contribution of the core. For the core rotation ($r<0.25 R_{\odot}$), we used Gaussian profiles that fit the different heliosesimic solutions proposed in \citep{1999MNRAS.308..405C,1998ESASP.418..329R,2017A&A...604A..40F} or we set a constant rotation at 435 nHz (see Table 1).
For each value of the Sun angular momentum, an INPOP adjustment is done and  $J^{\odot}_{2}$ is fitted. The $J^{\odot}_{2}$ obtained with INPOP20a in considering the Sun angular momentum from helioseismological measurements \citep{1998MNRAS.297L..76P} is given on the first line of Table \ref{tab:Sun}. This value,  (2.21 $\pm$ 0.01)$\times 10^{-7}$, is very close from the values deduced from SOHO (2.22 $\pm$ 0.009)$\times 10^{-7}$ and GONG (2.18 $\pm$ 0.005)$\times 10^{-7}$ \citep{2008A&A...477..657A}. 
It is also in good agreement with the previous analysis of the same data made by \citep{1998MNRAS.297L..76P} giving as an average estimate between GONG and SOHO, (2.18 $\pm$ 0.06) $\times 10^{-7}$. 
We emphasize that there is an important correlation (80 $\%$) between S and $J^{\odot}_{2}$  when both parameters are estimated in a global planetary fit.
Because of this high correlation, the Sun angular momentum S is not fitted in the INPOP adjustment instead we use the value from \citep{1998MNRAS.297L..76P}. 
The same choice has been made by \citep{2018NatCo...9..289G} who focused on using Messenger data for constraining Mercury and Earth orbits. Their obtained value of $J^{\odot}_{2}$ is also given in Table \ref{tab:Sun} and is consistent with our estimate as well as with \citep{1998MNRAS.297L..76P}. Using these different angular momenta to account for the Lense-Thirring effect in our global ephemerides fit, we extract the corresponding gravitational $J^{\odot}_{2}$, as shown in Tab. \ref{tab:Sun}. Our fitted values are in good agreement with those inferred from helioseismology \citep{1998MNRAS.297L..76P, 2008A&A...477..657A} with a value close to $2.2 \times 10^{-7}$. We emphasize that our value obtained using the very fast core (e.g. following \cite{2017A&A...604A..40F}) is smaller than the $J^{\odot}_{2} \approx 2.6 \times 10^{-7}$ found in \citet{2019ApJ...877...42S}, who used the same very fast solution. The reason of this difference is due to their large extent of the fast rotating core.  Our differences in $J^{\odot}_{2}$ coming from the different core rotations are much smaller than our error bars, which prevents us to disentangle these core rotations with the current planetary ephemerides. We will see that with the inclusion of the Bepi-Colombo simulations (BC), this conclusion could be different (Sect. \ref{sec:J2bc}).

\begin{table}
\caption{Sun Angular Momentum and oblateness. Is given in Column 3, the values of Sun $J^{\odot}_{2}$ obtained after fit using the values of the amplitude of the angular momentum given in Column 2. Different models of rotation (identified in Column 1) are used for estimating S. On the first line, are given the results obtained for INPOP20a. See Sect \ref{sec:srot} for the significance of the different rotation hypothesis. The final two columns give the results obtained after the introduction of BC in the fit sample. In particular in Column 5, we give the differences  $\Delta J^{\odot}_{2}$ between the reference  $J^{\odot}_{2}$ fitted using \citep{1998MNRAS.297L..76P} angular momentum value and the  $J^{\odot}_{2}$ obtained after fit using different values for the amplitude of the angular momentum given in Column 2. The uncertainties are given at 3-$\sigma$.}
\centering
\begin{tabular}{l l c c }
Type of rotation & S $\times 10^{48}$ & $J^{\odot}_{2}$  wo BC &  $\Delta J^{\odot}_{2} $ w BC\\
& g.cm$^{-2}$.s$^{-1}$& $\times 10^{7}$ & $\times 10^{10}$\\
\hline
INPOP20a  & 1.90 & 2.218 $\pm$ 0.03 &  0.0 $\pm$ 2.3\\
\citep{2018NatCo...9..289G}& 1.90  & 2.2710 $\pm$ 0.003 &\\
\\
Slow rotation &1.896 & 2.208 $\pm$ 0.03 &   -2 $\pm$ 2.3 \\
uniform rotation at 435 nHz & 1.926 & 2.210 $\pm$ 0.03 &   0 $\pm$ 2.3\\
Fast rotation &1.976 & 2.213 $\pm$ 0.03  &   2  $\pm$ 2.3 \\
Very fast rotation & 1.998 & 2.214 $\pm$ 0.03  &    3 $\pm$ 2.3\\
\hline
\end{tabular}
\label{tab:Sun}
\end{table}
 
\subsection{Bepi-Colombo simulations}

In using INPOP20a as a reference planetary ephemerides, we  simulate possible range bias  between Mercury and the Earth as they should be obtained by the radio science MORE experiment on board the Bepi-Colombo (BC) mission. Based on the assumption that the radio tracking in KaKa-Band keeps the 1~cm accuracy that has been monitored during the commissioning phase of the Bepi-Colombo mission in 2019 and 2020 \citep{2021SSRv..217...21I}, we suppose a daily acquisition of range tracking data \citep{thor20} during a period of 2.5 years, from 2026 to 2028.5. 
The simulated residuals obtained in using INPOP20a provide a reference against which can be tested the ephemerides integrated with non-GRT parameters (i.e., PPN $\beta \ne  \gamma \ne 1$ and $\dot{\mu}/\mu \ne 0$ or accounting for a massive graviton or a massless dilaton).
The capability of these alternative ephemerides to have a good fit with the GRT simulated observations tell us what constraints can be obtained on the PPN parameters.
In the same manner we have tested the sensitivity of the BC simulations to any change in the values of the Sun angular momentum (see Sect \ref{sec:J2bc}). 

\subsection{ WRSS filtering}

By construction, the planetary ephemerides cannot disentangle the contribution of the PPN parameters $\beta$, $\gamma$ and the Sun oblateness $J^{\odot}_{2}$ \citep{2015CeMDA.123..325F}. The introduction of the Lense-Thirring effect helps for individualizing the signature induced by PPN $\gamma$ but correlations between these parameters stay high. This is the reason why a direct adjustment of these three parameters together in a global fit leads to highly correlated determinations and under-estimated uncertainties. 
The detailed method of the Monte Carlo simulations  for building  the Weighted Residual Sum of Squares distribution is explained in \cite{2021arXiv211104499F}. 
It is based on the study of the instrumental noise variability of the INPOP adjustment and its impact on the WRSS estimation following the equation 
\begin{equation}
WRSS  = \frac{1}{N} \sum_{i=1}^{N} \frac{((O-C)_{i})^{2}}{\sigma_{i}^{2}}
\label{eq:chi2}
 \end{equation}
 where $(O-C)_{i}$ is the difference between the observation O and the observable C computed with INPOP (postfit residual) for the observation $i$, $\sigma_{i}$ is the {\it{a priori}} instrumental uncertainty of the observation $i$ and $N$ is the number of observations. We obtain an experimental WRSS  distribution from which we can estimate the probability of a postfit WRSS  to be explained by the instrumental uncertainties. We derive the quantiles corresponding to the 3-$\sigma$ of the WRSS  distribution  and we can estimate a confidence interval $[$WRSS $_{max}$:WRSS $_{min}]$ that contains 99.7$\%$ of the distribution. We used these WRSS $_{max}$ and WRSS $_{min}$ as thresholds for the selection of alternative ephemerides (estimated with non-GRT parameters) compatible at 99.7$\%$ with the observations.
 We proceed in the same manner with the Bepi-Colombo (BC) simulations. We add BC simulations to the INPOP20a data sample in radomly sampling Mercury-Earth range bias with 1$~$centimeter  instrumental uncertainty.

\section{Results}

\subsection{PPN tests without Bepi-Colombo}
\label{section2}

\begin{table}
\caption{PPN $\beta$, $\gamma$, $\dot{\mu}/\mu $ and $J^{\odot}_{2}$ confidence intervals given at 99.7 $\%$ of the reference WRSS  distribution, with and without Bepi-Colombo simulations. The first column indicates the method being used for obtaining the results given in Columns 4 (with INPOP20a) and 6 (with INPOP20a and BC simulations). LS stands for Least squares estimations.}
\centering
\begin{tabular}{l | c c c c}
& \multicolumn{1}{c}{$(\beta - 1)$} & \multicolumn{1}{c}{$(\gamma - 1)$} &   \multicolumn{1}{c}{$\dot{\mu}/\mu $}& $J^{\odot}_{2}$\\
& \multicolumn{1}{c}{$\times 10^{5}$} & \multicolumn{1}{c}{$\times 10^{5}$} &  \multicolumn{1}{c}{$\times 10^{13}$ yr$^{-1}$} & $\times 10^{7}$ \\
\hline
3-$\sigma$ WRSS  INPOP20a &  -1.12 $\pm$ 7.16  &   -1.69 $\pm$ 7.49  &    -1.03 $\pm$ 2.28 & 2.206 $\pm$ 0.03\\
non-GRT LS INPOP20a  &   -1.9 $\pm$  6.28 &    2.64 $\pm$  3.44  &   -0.88 $\pm$  0.29 & 2.165 $\pm$ 0.12\\
\\
3-$\sigma$ WRSS  INPOP20a + BC &  0.32 $\pm$ 5.00 &  0.09 $\pm$ 0.40 &   -0.19 $\pm$ 0.19  & 2.206 $\pm$ 0.009 \\
non-GRT INPOP20a + BC  &   $\pm$  1.06 &    $\pm$  0.23 &  $\pm$  0.01  &  $\pm$ 0.013\\
\hline
\end{tabular}
\label{tab:sres}
\end{table}

About 6000 runs are estimated with values of PPN $\beta$ and $\gamma$ different from unity and $\dot{\mu}/\mu$ different from 0.
 For each of these runs, alternate planetary ephemerides are integrated and fitted over the same data sample as INPOP20a in an iterative process. When the iterations converged, the obtained WRSS  is compared to the INPOP20a confidence interval. Only 23$\%$ of the runs with  $\beta$ and $\gamma$ different from unity have WRSS encompassed in this confidence interval and 60$\%$ with  non-zero $\dot{\mu}/\mu$.
 
Table  \ref{tab:sres} gives the deduced intervals for the three quantities and the Sun oblateness based on the WRSS  filtering. These intervals are all compatible with GRT and can be compared with the one obtained by direct least square fit (non-GRT LS), based on the INPOP20a data sample.

The non-GRT LS are obtained by adding ($\beta$, $\gamma$, $\dot{\mu}/\mu$) to the full INPOP20a adjustment together with the 402 other parameters (including the mass of the Sun and its oblateness, constrained by helio-seismology values). As for the WRSS  filtering, these estimations are also consistent with GRT. As a first consequence of important correlations induced between these parameters, $J^{\odot}_{2}$ and planet semi-major axis, in the global fit including $\beta$ and $\gamma$, the LS uncertainties are under-estimated and the WRSS  intervals are systematically larger than the LS results. 
\\

\subsection{PPN tests with Bepi-Colombo}
\label{sec:J2bc}

\begin{figure}
    \centering
    \includegraphics[width=9.5cm,height=9cm]{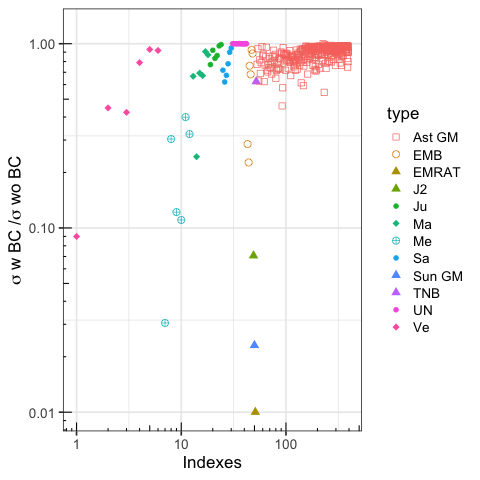}
    \caption{Distribution of the ratio between the parameter uncertainties obtained with and without BC simulations in log scale.The colors and shapes indicate the different types of parameters considered in the INPOP adjustment : Me,V,Ma,Ju,Sa,UN, EMB represent the ratio of the uncertainties for the 6 orbital initial conditions for Mercury, Venus, Mars, Jupiter, Saturn, Uranus, Neptune and the Earth-Moon barycenter respectively. $J^{\odot}_{2}$, Sun GM, EMRAT and TNB give the ratios for the Sun oblateness and mass, the ratio between the Earth and the Moon masses and the mass of the TNO ring respectively. Finally Ast${_GM}$ indicates the ratio for the 343 Main Belt asteroid masses.}
    \label{fig:sigcdi}
\end{figure}

The first aspect to consider when one introduces BC simulations in planetary adjustments is the impact on the determination of planetary orbits by the means of the evolution of the covariance matrix of the planetary orbit initial conditions and other parameters of the fit.
On Fig. \ref{fig:sigcdi}, we plotted the ratio between the standard deviations (defined as the square root of the diagonal terms of the covariance matrix deduced from the least squares adjustment) for the 402 parameters of INPOP20a in GRT obtained in including the BC simulation ($\sigma$ w BC) and  without the BC simulation ($\sigma$ wo BC). As one can see on this figure, the highly improved parameters, besides the Mercury and the Earth-Moon barycenter orbits,  are the Earth-Moon mass ratio, the mass of the Sun and its oblateness. 
As secondary perturbers of the Mercury-Earth distance, Venus sees also its orbit improved as well as Mars. At a lower level, Jupiter and Saturn orbits are also better estimated when the Main Belt asteroid masses are marginally improved. This indicates a slight reduction of the mass uncertainties for some of the perturbers.

In GRT, the results of the $J^{\odot}_{2}$ LS adjustment including BC simulations for different models of Sun core rotations are given in the Column 4 of the Table \ref{tab:Sun}. 
One can notice a significant reduction of the 3-$\sigma$ LS uncertainty from 3$\times 10^{-9}$ with INPOP20a to 2 $\times 10^{-10}$ when including the BC simulations (as noticeable in Fig. \ref{fig:sigcdi} as well). At this level of accuracy, the differences between  Sun core rotation modes appear to be detectable . 
More precisely,  despite the fact that the estimated values for the Sun oblateness are still consistent at 2$\sigma$ with  \citep{2008A&A...477..657A}, they differ from one Sun core rotation mode to another by a maximum of 5 $\times 10^{-10}$ (between the slow and the very fast rotations) , which is more than 2 times bigger than the 3-$\sigma$ LS uncertainty.  This can indicate the possible disentangling of the different (uncertain) solution of helioseismology in the core  thanks to the addition of the BC data. If we consider the direct fit of $J^{\odot}_{2}$ together with the non-GRT parameters or the WRSS $J^{\odot}_{2}$  determination given in the last columns of Table \ref{tab:sres}, the obtained 3-$\sigma$ LS or WRSS uncertainties are improved relative to the solutions without BC, but they remain  larger than the uncertainty obtained in GRT. 
In this context, the detection of the different models for the Sun core rotation appears then to be out of reach when we consider a simultaneous estimation of non-GRT parameters and $J^{\odot}_{2}$ but detectable in a direct LS estimation in GRT.

On Table \ref{tab:sres} are also given the results obtained by adding the BC simulations to the INPOP20a data sample for PPN parameters. A first striking result is  that BC will improve drastically the constraint on the possible violation of GRT through the PPN parameters $\beta$ and $\gamma$. For the 3-$\sigma$ WRSS  filtering, the most spectacular is the estimation of the $\gamma$ parameter which gains a factor 19 in comparison with the INPOP20a results. The constraint on $\beta$ is less improved, of about a factor 1.5. We also note an improvement of the LS results with and without BC of about a factor 6 for $\beta$ and 15 for $\gamma$. 
For $\dot{\mu}/\mu$ the introduction of the BC simulations induces an improvement of about a factor 12 with the WRSS filtering and 30 in the direct LS. With such a constraint, in the perspective of measuring $\dot{G}/G$, it will be important to have independent and accurate constraints for the Sun mass loss which has currently a higher uncertainty ($0.61 \times 10^{-13}$ yr$^{-1}$).

\subsection{The massive Graviton}
\label{sec:graviton}

In following \cite{2020PhRvD.102b1501B,bernus2019}, the hypothesis of a massive graviton is introduced in INPOP in considering an additional acceleration $ a_{grav}$ given by with the equation :
\begin{equation}
\delta a_{grav} = \frac{1}{2{\lg^2}}\sum_{B\ne A} GM_B \frac{\bm{x}_A - \bm{x}_B}{|\bm{x}_A - \bm{x}_B|} + O(\lg^{-3}) 
\label{eq:graviton}
\end{equation}
where  $\lg$ is the Comptom length, related to the mass of the graviton $m_{g}$ by $h /(c \times {m_{g}})$.
Alternative ephemerides are integrated including $ a_{grav}$ and fitted to the INPOP22a data sample with and without Bepi-Colombo. A selection of acceptable alternative ephemerides is done following the WRSS filtering and a limit in the maximum of the Compton length value is obtained and given in Table \ref{tab:graviton}.  Two regimes of detection are possible for the mass of the graviton: the dynamical mode with the detections of gravitational waves and the static mode with the interaction of the mass of the static graviton with planetary orbits.  With the first round of analysis of the Virgo-Ligo catalog (GWTC-1 \cite{PhysRevX.9.031040}), the Compton length  measured by gravitational waves was closed from the obtained with INPOP19a  \cite{2020PhRvD.102b1501B}. But the accuracy of the third version of the Virlo-Ligo catalog GWTC-3 \cite{GWTC3}  induces a better constraint on the  graviton mass compared to the present INPOP measure (i.e. in Table  \ref{tab:graviton}, GWTC-3 versus INPOP22a wo BC). However, the addition of the BC simulation seems to push the INPOP result at the level of GWTC-3.  We also note that the results obtained with the WRSS filtering give a limit compatible with the one obtained by \cite{2020CQGra..37i5007D} with a covariance analysis including independent BC simulations .
Interpretations in terms of Yukawa potential constraint are also given in Table   \ref{tab:graviton} following the potential approximation $V(r)=V_{Newton}(r)(1+\alpha\mathrm{exp}(-r/\lambda))$  with $r / \lambda_{g} << 1$ and $\alpha < 1$

\begin{table*}
\caption{Limits obtained for the Compton length $\lg$ in km as defined in Eq.  \ref{eq:graviton}. Are also given the corresponding values in term graviton mass $m_g$ in eV$/c^2$.  Are also indicated for comparisons the values obtained with INPOP19a \citep{2020PhRvD.102b1501B} and \cite{2020CQGra..37i5007D} as well as the estimations for the dynamical mode from Virgo-Ligo  GWTC-1 \cite{PhysRevX.9.031040} and GWTC-3 \cite{GWTC3}}
\centering
     \begin{tabular}{c | l l | c c c | c}
&\multicolumn{2}{c |}{GW}& \multicolumn{1}{c}{INPOP19a } & \multicolumn{2}{c}{INPOP22a}& \cite{2020CQGra..37i5007D} \\
&{TC-1  \cite{PhysRevX.9.031040} } & {TC-3 \cite{GWTC3}}& \multicolumn{1}{c}{\citep{2020PhRvD.102b1501B}} & {wo BC} &{with BC}& {with BC}\\
CL& {0.90}    & {0.90}                  & 0.90   & {0.90}  & {0.90}  &\\
    \hline
       Yukawa suppression                 &     &                   &                                                & &&\\
         $\lg$ $\times10^{-13}$   [km]            &                {2.6} &  {9.77}  &  3.93    & {5.50} & {8.32} & 8.8 \\
         $m_g$   $\times10^{23}$   [eV$/c^2$]     & {4.7} & {1.27}  & 3.16   &   {2.17} &  {1.51} &\\ 
    \hline
         Fifth Force                           &                   &           &                                        &&&\\
    $\lambda/\sqrt{|\alpha|}$ for $\alpha > 0$ [km] &         &                            &  3.93     &{5.50}& {8.32}&\\
    $\lambda/\sqrt{|\alpha|}$ for  $\alpha < 0$ [km] &         &                         &3.77  & &&\\
    \hline
    \hline
    \end{tabular}
        \label{tab:graviton}

\end{table*}

\subsection{The massless dilaton in linear coupling}
\label{sec:dilaton}
As in \cite{bernus2022}, we introduce the modified equation of motion for the planetary and lunar ephemerides  accounting for three matter-field couplings depending on the nature of the bodies and derived from the theory of the massless dilaton firtly proposed by \cite{damour2010prd} and then adapted to planetary ephemerides by \cite{minazzoli2016prd} and \cite{bernus2022}. These three parameters are $\alpha_0$, the universal coupling, $\alpha_G$ for the gazeous planets and $\alpha_T$ for the telluric planets. The same method as for the PPN parameters and the massive graviton is used. We first sample randomly the dilaton parameters introduced in INPOP ($\alpha_0$,$\alpha_T$,$\alpha_G$). For each tuple, we then built a planetary and lunar ephemeris by integration of the modified EIH and fit to the observations (INPOP22a with and without Bepi-Colombo simulations). A selection based on the WRSS value is done. The distributions of the selected ephemerides in terms of  ($\alpha_0$,$\alpha_T$,$\alpha_G$) are given in Table \ref{tab:dilaton}.
The improvement brought by the Bepi-Colombo simulations is clearly visible, especially for the telluric constraint $\alpha_T$ for which the intervalle of possible values is divided by a factor about 10. For $\alpha_0$, the improvement is about a factor 2 and about factor 1.5 for $\alpha_G$. Are also given the deduced constraint on $\gamma$ as $\gamma=\frac{1- {\alpha_0^2}}{1+ {\alpha_0^2}}$


	\begin{table}\caption{Intervals of possible values for the 3 dilaton parameters: $\alpha_0$, the universal coupling,  $\alpha_T$ the telluric planet coupling and $\alpha_G$ the gazeous planet coupling. For comparisons are also given the values obtained by \cite{bernus2022} with INPOP19a. }
\centering
		\begin{tabular}{lrr || c }
		&   \multicolumn{2}{c}{INPOP19a \cite{bernus2022}} & INPOP22a with BC  \\
		& \multicolumn{2}{c}{} & this work  \\
			 Confidence:      & 90\%            & 99.5\%   &  99.5\% \\
			\hline
			$\alpha_0(\times 10^{5})$  & $-0.94\pm 5.35$ &$1.01\pm23.7 $  &{$\pm 11.29 $}\\
			$\alpha_T(\times 10^{6})$  & $0.24\pm1.62$    & $0.00\pm24.5 $ &{$\pm 2.80 $}\\
			$\alpha_G(\times 10^{5})$  & $0.01\pm4.38$   & $-1.46\pm12.0 $  & {$\pm 8.29 $}\\
			\hline
			$(\gamma-1) \times 10^{{8}}$ & 0.2 $\pm$ 6  & 0.2 $\pm$ 11.2 &  {$\pm  2.5$}\\
			\hline
					\end{tabular}
					        \label{tab:dilaton}
		\end{table}

\section{Conclusions}
New constraints on PPN parameters $\beta$,$\gamma$, possible time variations of the Sun gravitation mass are given using the latest versions of the INPOP planetary ephemerides and WRSS filtering. Lense-Thirring acceleration has been added to the INPOP equations of motion opening the door to the detection of the Sun core rotation mode, especially in the frame of the Bepi-Colombo mission. The first Mercury flyby by BC occurred in October 2021 and the radio science data obtained during this first flyby are currently analysed by the MORE team. They will be included in INPOP soon, allowing new GRT tests and measures of the mass of the graviton  and the three coupling parameters of the massless dilaton. These estimations will be compared with the simulations proposed here.

\section*{Acknowledgments}
The authors thank the French Space Agency (CNES)  as well as the Programme National PNGRAM for their long term support. This work also benefits from the Observatoire de la Côte d'Azur fundings.

%


\end{document}